\documentclass[preprint]{elsarticle}
\usepackage{graphicx}
\usepackage{epsfig}
\usepackage{amsfonts}
\usepackage{amssymb}
\usepackage{amsmath}
\usepackage{natbib}
\title{Discrete spectrum of kink velocities in Josephson structures: the nonlocal double sine-Gordon model}
\author[miet]{G.L. Alfimov\corref{cor1}}
\ead{galfimov@yahoo.com}

\author[fian]{A.S. Malishevskii}
\ead{malish@lebedev.ru}

\author[miet]{E.V. Medvedeva}
\ead{elinamedvedeva87@gmail.com}

\cortext[cor1]{Corresponding author}
\address[miet]{National Research University of Electronic Technology, Moscow
124498, Russia}

\address[fian]{P.~N.~Lebedev Physical Institute of the Russian Academy of Sciences, Moscow 119991, Russia} \ead{malish@lebedev.ru}

\begin{document}
\begin{frontmatter}

\begin{abstract}
We study a model of Josephson layered structure which is characterized
by two peculiarities: (i) superconducting layers are thin; (ii) due to
suppression of superconducting states in superconducting layers the
current-phase relation is non-sinusoidal and is described by two sine
harmonics. The governing equation is a nonlocal generalization of
double sine-Gordon (NDSG) equation. We argue that the dynamics of
fluxons in the NDSG model is unusual. Specifically, we show that there
exists a set of particular velocities for non-radiating fluxon
propagation. In dynamics the presence of these ``priveleged''
velocitied results in phenomenon of {\it quantization of fluxon
velocities}: in our numerical experiments a travelling kink-like
excitation radiates energy and slows down to one of these particular
velocities, taking a shape of predicted $2\pi$-kink. This situation
differs from both, double sine-Gordon local model and the nonlocal
sine-Gordon model, considered before. We conjecture that the set of
these velocities is {\it infinite} and present an asymptotic formula
for them.

\end{abstract}

\end{frontmatter}

\section{Introduction}
Since mid-60's it is known that the description of a long contact
between two superconductors (Josephson junction, JJ) is based on the
classical sine-Gordon equation:
\begin{eqnarray}
\sin\varphi+\omega_J^{-2}\varphi_{tt}=\lambda_J^2\varphi_{xx}.
\label{sG}
\end{eqnarray}
Here $\varphi=\varphi(x,t)$ is the phase difference of the order
parameters in the superconducting banks, $\omega_J$ is the Josephson
plasma frequency, $\lambda_J$ is the Josephson length. The first term
in the left-hand side of Eq.~(\ref{sG}) comes from the formula for
supercurrent across the JJ
\begin{eqnarray}
J(\varphi)=J_c\sin\varphi, \label{J-sin}
\end{eqnarray}
where $J_c$ is the critical Josephson current.

In mid-70's it was recognized that in some situations the second
derivative term in the right-hand side of Eq.~(\ref{sG}) should be
corrected. In particular, this should be done if the London penetration
depth $\lambda_L$ becomes comparable with Josephson length $\lambda_J$.
Then Eq.~(\ref{sG}) must be replaced by integral equation of the type
\begin{eqnarray}
\sin\varphi(x,t)+\omega_J^{-2}\varphi_{tt}(x,t)=\lambda_J^2
\frac{\partial}{\partial x}\int dx' G(x,x')\varphi_{x'}(x',t).
\label{eq-G}
\end{eqnarray}
In literature Eq.~(\ref{eq-G}) has been called {\it nonlocal
sine-Gordon equation} \cite{AAM}. Explicit form of the kernel $G(x,x')$
depends on physical and geometrical properties of JJ. Some survey of
recent results in this field (sometimes called {\it nonlocal Josephson
electrodynamics}) including a list of kernels which have been used in
literature can be found in \cite{AAM}. It has been found that dynamics
of vortex in nonlocal Josephson electrodynamics has essential
peculiarities. One of the most exciting of them is forming of bound
states of more then one flux quanta and the quantization of velocities
of these structures. This issue has been discussed in many papers
\cite{AEKM,AEL,MSU99,MSU-2000,MSU-10pi}
\begin{figure}
\centerline{\includegraphics[scale=0.6]{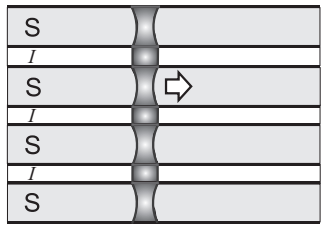}}
 \caption{Layered structure: S --- superconducting layers, I --- tunnel layers. A vortex formation moves from left to right.
  }\label{struct}
\end{figure}

The principles of nonlocal Josephson electrodynamics can be also
applied to stacked Josephson structures. In particular, in
\cite{AOSU} it has been shown that structures consisting of
alternating flat superconducting and tunnel layers (see
Fig.~\ref{struct}) can be described by Eq.~(\ref{eq-G}) with the
Kac-Baker kernel $G(x,x')\sim e^{-\gamma|x-x'|}$,
\begin{eqnarray}
\sin\varphi(x,t)+\omega_J^{-2}\varphi_{tt}(x,t)= \Lambda
\frac{\partial}{\partial x}\int dx' {\rm
e}^{-|x-x'|/\lambda_{eff}}\varphi_{x'}(x',t), \label{eq-exp-dimension}
\end{eqnarray}
where
\begin{eqnarray*}
\Lambda=\frac{(\lambda_L+d)\lambda_J^2}{2\lambda_L\sqrt{(L+d)L}},
\end{eqnarray*}
$\lambda_L$ is the London penetration depth, $2d$ is the thickness of
the each tunnel layer, $2L$ is the thickness of the each
superconducting layer and
\begin{eqnarray*}
\lambda_{eff}\equiv\lambda_L \sqrt{\frac{L}{L+d}}.
\end{eqnarray*}
In \cite{AOSU} Eq.~(\ref{eq-exp-dimension}) has been derived under the
following assumptions: (i) superconducting layers are identical, (ii)
tunnel layers are identical, (iii) vortex formation is symmetric, i.e.
the phase difference and the magnetic field in all the JJs are
identical, and (iv) superconducting layers are thin. The model
(\ref{eq-exp-dimension}) possesses many fascinating features
\cite{AEKM}. In particular, it allows to describe travelling vortices
of more then one flux quantum. At the same time, the mobility of
traditional fluxons in this model is essentially reduced.

Another motivation to modify Eq.~(\ref{sG}) is caused by {\it
non-sinusoidal character of current-phase relation.} Generally
speaking, Eq.~(\ref{J-sin}) represents the first term of the
relation \cite{Likharev,TK}
\begin{eqnarray}
J(\varphi)=J_c\sin\varphi+J_2\sin 2\varphi+J_3\sin 3\varphi+\dots
\label{J-sum}
\end{eqnarray}
In many situations $J_2\gg J_k$, $k=3, 4\ldots$. Therefore in many
studies (see e.g. \cite{GKKB07,Komis-08,Dubna,Bakur-13}) the
current-phase relation has been assumed to be of the form
\begin{eqnarray}
J(\varphi)=J_c \sin\varphi+ J_2 \sin 2\varphi. \label{DoubleHarm}
\end{eqnarray}
The sign of $J_2$ depends on a mechanism of suppression of
superconducting state in electrodes. In SIS-type junctions $J_2>0$ due
to suppression of superconductivity near the tunnel barrier by a
supercurrent \cite{Kupr-92,Gol-2005}. In SNINS and SFIFS junctions
there exists another mechanism, associated with proximity effect
\cite{Kupr-92,Gol-2005,Gol-2004} which may result in negative values of
$J_2$.

The equation for the phase difference $\varphi$ in the case of current-phase
relation (\ref{DoubleHarm}) reads
\begin{eqnarray}
\sin\varphi(x,t)+2A\sin
2\varphi(x,t)+\omega_J^{-2}\varphi_{tt}(x,t)=\lambda_J^2\varphi_{xx}(x,t),
\label{sG2}
\end{eqnarray}
where $A\equiv J_2/2J_c$. This equation is {\it the double sine-Gordon
equation} which has been widely discussed in both physical and
mathematical literature.

In this paper we study an effect of non-sinusoidal current-phase
relation on mobility of fluxons in Josephson structures with nonlocal
electrodynamics. We present a model of infinite Josephson structure
consisting of alternating superconducting and tunnel layers. It will be
assumed that current-phase relation is of the form (\ref{DoubleHarm}).
Also we assume  that S-layers are thin, and replace the second
derivative term in Eq.~(\ref{sG}) by nonlocal term. Repeating the
reasoning of~\cite{AOSU} (the details of derivation can be found in
\ref{Deriv}) we arrive at the equation
\begin{eqnarray}
\sin\varphi(x,t)+2A \sin 2\varphi+\omega_J^{-2}\varphi_{tt}=
\Lambda\frac{\partial}{\partial x}\int dx' {\rm
e}^{-|x-x'|/\lambda_{eff}}\varphi_{x'}(x',t).
\label{eq-exp-2-dimension}
\end{eqnarray}

The main output of our study can be formulated as follows. The
properties of free propagation of Josephson vortices in the model
(\ref{eq-exp-2-dimension}) {\it essentially differ}  from ones in (i)
the traditional sine-Gordon model, (ii)  double sine-Gordon local
model, Eq.~(\ref{sG2}), and (iii) nonlocal sine-Gordon model,
Eq.~(\ref{eq-exp-dimension}). Specifically, it has been found that
there exist {\it ``privileged''} velocities of free fluxon propagation
in radiationless regime. Contrary to the traditional sine-Gordon and
double sine-Gordon equations, the set of these velocities is {\it
discrete} and {\it infinite}. The nonlocal sine-Gordon equation
possesses similar property, but for the vortices of topological charge
 greater than 1 only (these entities can be regarded as ``bound
states'' of simple fluxons). It is worth mentioning that in the case of
nonlocal double sine-Gordon model the shapes of fluxons corresponding
to different ``priveleged'' velocities are {\it nearly the same}, the
difference takes place in the asymptotics of the ``tails'' of these
vortices.

The paper is organized as follows. In Section 2 we transform~Eq.~(\ref{eq-exp-2-dimension}) into a dimensionless equation which
depends on two external parameters, $\lambda$ and $A$. We call it {\it
nonlocal double sine-Gordon equation} and discuss some its features in
general. In Section 3 we give arguments for {\it quantization} of
fluxon velocities described by this model. Specifically, we present the
dependencies of these velocities $v_n(\lambda)$, $n=0,1,\ldots$, on the
parameter $\lambda$  and give a formula for asymptotics of these
dependencies. In Section 4 we report on results of numerical simulation
of fluxon evolution. We show that the dynamics ``feels'' the velocities
of this spectrum. Section 5 contains summary and discussion. Some
physical details, including the derivation of the basic nonlocal
equation~(\ref{eq-exp-2-dimension}), are postponed in Appendix A.

\section{Preliminaries}\label{prelim}
\subsection{Dimensionless form}\label{DimLess}
 In the dimensionless variables,
\begin{eqnarray*}
\zeta\equiv \sqrt{\frac{L+d}{\lambda_L+d}}\frac{x}{\lambda_J},\quad \tau
\equiv \omega_J t
\end{eqnarray*}
Eq.~(\ref{eq-exp-2-dimension}) takes the form
\begin{eqnarray}
\sin \varphi+2A \sin 2\varphi+\varphi_{\tau \tau}=
\frac{1}{2\lambda}\frac{\partial}{\partial \zeta}\int d\zeta' {\rm
e}^{-|\zeta-\zeta'|/\lambda} \varphi_{\zeta'}(\zeta',\tau),
\label{eq-exp-2}
\end{eqnarray}
where $\varphi\equiv \varphi(\zeta,\tau)$ and
\begin{eqnarray*}
\lambda\equiv
\frac{\lambda_{eff}}{\lambda_J}\sqrt{\frac{L+d}{\lambda_L+d}}=\frac{\sqrt{L}\lambda_L}{\lambda_J\sqrt{\lambda_L+d}}.
\end{eqnarray*}
Eq.~(\ref{eq-exp-2}) has the energy integral, $d{\cal W}/d\tau=0$,
where
\begin{eqnarray}
{\cal W}&=&\int~d\zeta\left[1-\cos\varphi+A(1-\cos
2\varphi)+\frac12\left(\frac{\partial \varphi}{\partial
\tau}\right)^2\right]+\nonumber\\[2mm]
&+&\frac{1}{4\lambda}\int\int~d\zeta_1~d\zeta_2\exp\left\{-\frac{|\zeta_1-\zeta_2|}{\lambda}\right\}\frac{\partial\varphi(\zeta_1,\tau)}{\partial
\zeta_1}\frac{\partial\varphi(\zeta_2,\tau)}{\partial \zeta
_2}.\label{FinalE}
\end{eqnarray}
This follows from the formula (\ref{EnIntegral})  (see \ref{Deriv}),
which can be written as
\begin{eqnarray*}
W=\frac{\phi_0 j_c\lambda_J}{2\pi
c}\sqrt{\frac{\lambda_L+d}{L+d}}{\cal W}.
\end{eqnarray*}
We call Eq.~(\ref{eq-exp-2}) {\it nonlocal double sine-Gordon
equation} (NDSG).\medskip

\subsection{Kink solutions}\label{KinkSol}
We consider travelling wave solutions,
$\varphi(\xi)=\varphi(\zeta-v\tau)$, of Eq.~(\ref{eq-exp-2}) which
satisfy the equation
\begin{eqnarray}
\sin \varphi+2A \sin 2\varphi+v^2\varphi_{\xi \xi}=
\frac{1}{2\lambda}\frac{d\,}{d \xi}\int  {\rm
e}^{-|\xi-\xi'|/\lambda} \varphi_{\xi'}(\xi')~d\xi'. \label{TrW}
\end{eqnarray}
A single Josephson vortex (fluxon) corresponds to $2\pi$-kink solution
of Eq.~(\ref{TrW}). It obeys the boundary conditions
\begin{eqnarray*}
\lim_{\xi\to-\infty}\varphi(\xi)=0, \quad
\lim_{\xi\to+\infty}\varphi(\xi)=2\pi.
\end{eqnarray*}\medskip

\subsection{Approximations}\label{Approx}
The operator ${\cal L}$ defined as
\begin{eqnarray*}
{\cal L}\varphi=\frac{1}{2\lambda}\frac{d}{d \xi}\int {\rm
e}^{-|\xi-\xi'|/\lambda} \varphi_{\xi'}(\xi')~d\xi'
\end{eqnarray*}
is a Fourier multiplier operator, $\widehat{{\cal
L}\varphi}(k)=\widehat{\cal L}(k)\cdot\widehat{\varphi}(k)$, where
$\widehat{f}(k)$ means the Fourier transform of function $f(\xi)$
and $\widehat{\cal L}(k)$ is the symbol of operator ${\cal L}$,
\begin{eqnarray}
\widehat{\cal L}(k)=-\frac{k^2}{1+k^2\lambda^2}.\label{Symb}
\end{eqnarray}
In the limit $\lambda\ll 1$ one can replace the symbol
$\widehat{\cal L}(k)$ by its Taylor approximations. One term
approximation, $\widehat{\cal L}(k)\approx-k^2$, returns us to the
double sine-Gordon case
\begin{eqnarray}
\sin \varphi+2A \sin 2\varphi=(1-v^2)\varphi_{\xi \xi}.
\label{2SGTrW}
\end{eqnarray}
For $A > -1/4$ and $v^2<1$ Eq.~(\ref{2SGTrW}) admits exact
$2\pi$-kink solution
\begin{eqnarray}
\tilde{\varphi}(\xi)=\pi + 2\arctan\left(\frac1{\sqrt{1+4A}}
\sinh\left(\frac{\sqrt{1+4A}}{\sqrt{1-v^2}}\xi\right)\right).
\label{2SG_Sol}
\end{eqnarray}
Simple phase plane analysis shows that this $2\pi$-kink solution is
unique. Therefore, in the local limit the model allows for fluxons
which can travel with arbitrary velocity $v^2<1$.

Two term approximation for (\ref{Symb}) reads
\begin{eqnarray*}
\widehat{\cal L}(k)\approx-k^2+\lambda^2 k^4.
\end{eqnarray*}
In this case $2\pi$-kink solution obeys the 4-th order ODE
\begin{eqnarray}
\sin \varphi+2A \sin 2\varphi+(1-v^2)\varphi_{\xi
\xi}+\lambda^2\varphi_{\xi\xi\xi\xi}=0.\label{4thOrd}
\end{eqnarray}
After scaling of independent variable, $\eta=\xi/\sqrt{1-v^2}$,
one arrives at the equation
\begin{eqnarray}
\sin \varphi+2A \sin 2\varphi+\varphi_{\eta
\eta}+\delta^2\varphi_{\eta \eta\eta \eta}=0,\quad
\delta=\frac{\lambda}{\sqrt{1-v^2}}.\label{4thOrdScale}
\end{eqnarray}
The fourth-order term in (\ref{4thOrd}) becomes essential for
$\lambda\sim\sqrt{1-v^2}\ll1$.

\subsection{Embedded solitons} \label{Embedded}

Let $\lambda\ne 0$. Linearizing Eq.~(\ref{TrW}) near the
equilibrium $\varphi=0$ and seeking for small amplitude
excitations $\varphi\propto e^{ik\xi}$ we arrive at the equation for
$k$
\begin{eqnarray}
-v^2k^2+\frac{k^2}{1+\lambda^2k^2} = 1 + 4A . \label{DispRel}
\end{eqnarray}
Direct calculation shows that if $A>-1/4$ and $v^2<1$
Eq.~(\ref{DispRel}) has a single pair of real roots $k=\pm
k_0(\lambda)$. In terminology of \cite{ES1Gen,ES2Gen}, we are in
situation when the resonance prohibits propagation of localized wave
for Eq.~(\ref{eq-exp-2}) and so called {\it embedded solitons} may
appear. In this case, the velocity $v$ of embedded soliton (i.e.
$2\pi$-kink), {\it is not arbitrary} but should be ``adjusted'' to
avoid oscillatory asymptotics of its tail due to merging with linear
modes. Typically, each value $v$ is isolated and belongs to some
discrete set. This set may be empty (i.e. no localized waves propagate)
or include either finite or infinite number of values. Note that in the
case $\lambda=0$ no resonance occurs, since in this case
Eq.~(\ref{DispRel}) has no real roots for $v^2<1$.

\section{Kink solutions: results}\label{KinkSolRes}

\subsection{Numerical results for Eq.~(\ref{TrW})}\label{NR_Main}

In order to study $2\pi$-kink solutions for Eq.~(\ref{TrW}) we employ
an approach described in \cite{AEKM,AlfMed11}. Since
\begin{eqnarray*}
q(\xi)=\frac1{2\lambda}\int  {\rm e}^{-|\xi-\xi'|/\lambda}
\varphi_\xi(\xi')~d\xi'
\end{eqnarray*}
is a solution of equation $-\lambda^2q_{\xi\xi} + q = \varphi_\xi$, we
replace Eq.~(\ref{TrW}) with the system
\begin{eqnarray}
&&v^2\varphi_{\xi\xi} = q_\xi + \sin \varphi + 2A \sin 2\varphi,\label{Eq1} \\
&&-\lambda^2q_{\xi\xi} + q = \varphi_{\xi}, \label{Eq2}
\end{eqnarray}
which is completely equivalent to Eq.~(\ref{TrW}) \cite{NonlinDyn11}.
Then the study of $2\pi$-kinks for Eq.~(\ref{TrW}) may be reduced to
analysis of heteroclinic separatrices which connect equilibrium states
$\varphi=0$ and $\varphi=2\pi$ in 4D phase space of the system
(\ref{Eq1})~---~(\ref{Eq2}). Generically, these separatrices may exist
not for any choice of parameters $\lambda$ and $v$. For a fixed
$\lambda$, numerical computation of allowed velocities $v$ can be done
by seeking for zeroes of some function $R_\lambda(v)$ (technical
details can be found in \cite{AEKM,AlfMed11}). Theoretically, this
approach allows to find {\it all} possible kink velocities from a given
interval $[\tilde{v}_1,~\tilde{v}_2]$.

\begin{figure}
\centerline{\includegraphics[scale=0.5]{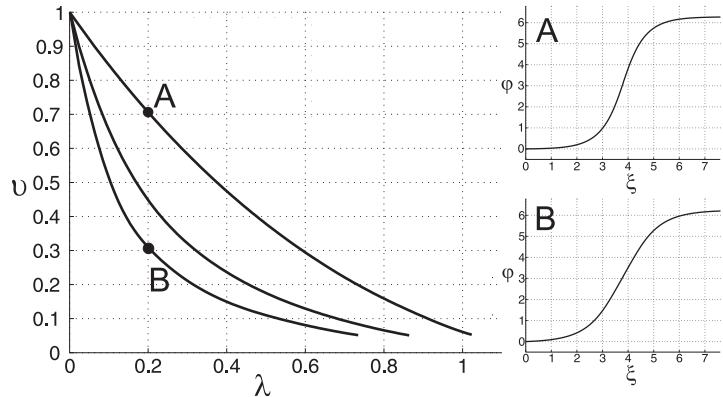}}
 \caption{Kink solutions of Eq.~(\ref{TrW}).
Values of $v_n$ versus $\lambda$ are shown for the first three
solutions. Profiles of the first and third kinks for $\lambda=0.2$ are
shown in the inserts. Both of them are indistinguishable from the
profile of the kink (\ref{2SG_Sol}), corresponding to the local case
$\lambda=0$.}\label{fig-example-2}
\end{figure}

Numerical study  confirms the existence of embedded solitons for
Eq.~(\ref{TrW}). Specifically, for a fixed $\lambda$ and $A>0$ there is
a discrete set of $2\pi$-kink solutions of Eq.~(\ref{eq-exp-2}). Each
of them corresponds to its own velocity $v_n$, $n=1,2,\ldots$. We
conjecture (see Sect.~\ref{Asym_v}) that the spectrum of velocities
{\it is infinite} and $v_n$ accumulate to zero velocity when $n\to
\infty$. The curves $v_n(\lambda)$ corresponding to the three highest
velocities $v_1$, $v_2$ and $v_3$ are shown in Fig.~\ref{fig-example-2}
(left panel). Two profiles of the $2\pi$-kinks corresponding to points
A and B ($\lambda=0.2$) are shown in right panel of
Fig.~\ref{fig-example-2}. It is worth noting that the {\it central
parts} of  kink profiles for different $v$  are very close to the
profile of kink (\ref{2SG_Sol}), corresponding to the local case
$\lambda=0$. The difference between them becomes substantial in the
asymptotics of their tails.

All the curves $v_n(\lambda)$ in the diagram of Fig.~\ref{fig-example-2} are originated at the point $\lambda=0$,
$v=1$. In vicinity of this point the dependence $v_n(\lambda)$ can
be described by Eq.~(\ref{4thOrdScale}).

\subsection{The limit case $\lambda\sim\sqrt{1-v^2}\ll1$}\label{l_ll_1}

If $\lambda\sim\sqrt{1-v^2}\ll1$ and
$\delta=\lambda/\sqrt{1-v^2}\sim 1$ the profile of the
kink is described by Eq.~(\ref{4thOrdScale}). The dispersion
relation for this case reads
\begin{eqnarray}
k^2+\delta^2k^4 = 1 + 4A. \label{DispRelOrd4}
\end{eqnarray}
This equation admits a pair of real roots $k=\pm \tilde{k}_0(\lambda)$
for $A>-1/4$. Following the ``embedded soliton'' approach, one can
expect that kink solution may exist  only for discrete set of values of
governing parameter $\delta$. Numerical study fulfilled for various
values of $A>0$ confirms the existence of this discrete set
$\delta_1,\delta_2,\ldots$  (see Table \ref{T2} in Sect.~\ref{Asym_v}).
Therefore, in a vicinity of the point $\lambda=0$, $v=1$ in the diagram
in Fig.~\ref{fig-example-2} the dependence $v_n(\lambda)$ for
$n=1,2,\ldots$ obeys the asymptotic formula
\begin{eqnarray*}
v_n(\lambda)\simeq\sqrt{1-\frac{\lambda^2}{\delta_n^2}}.
\end{eqnarray*}

\subsection{Asymptotic formula for $v(\lambda)$}\label{Asym_v}

Numerical study allows to suppose that for $\lambda$ fixed the values
of $v_n$ tend to zero as $n$ grows. Simultaneously, if $v$ is fixed,
the corresponding values of $\lambda_n$ also tend to zero when $n$
increases. To describe asymptotical properties of the set of families
$v_n(\lambda)$ one can apply the approach presented in \cite{arXiv}.
The main statement of \cite{arXiv} concerns the equation
\begin{eqnarray}
G_\lambda \varphi=F(\varphi),\label{MainEq}
\end{eqnarray}
where $\varphi(\xi)$ and $F(\varphi)$ are real-valued functions and
$\varphi(\xi)$ is defined on whole $\mathbb{R}$. Assume that
$G_\lambda$ is a Fourier multiplier operator which depends continuously
on real parameter $\lambda$ and $G_0=\Omega^2 d^2/d\xi^2$ where
$\Omega$ is a real number. The symbol ${\hat G}_\lambda(k)$ of the
operator $G_\lambda$ is supposed to be an even function.\medskip

{\bf Conjecture}, \cite{arXiv}. {\sl Assume that

(a)  $\varphi=\varphi_+$ and $\varphi=\varphi_-$ are zeroes of the
function $F(\varphi)$;

(b) $F'(\varphi_+)=F'(\varphi_-)>0$ and the equation
\begin{eqnarray}
{\hat G}_\lambda(k)-F'(\varphi_{1})=0,\label{DisRelKink}
\end{eqnarray}
has only one pair of real roots $k=\pm k(\lambda)$, $k(\lambda)>0$ and
$k(\lambda)\to\infty$ as $\lambda\to 0$;

(c) the equation $\Omega^2 d^2\varphi/d\xi^2=F(\varphi)$ has a kink
solution ${\tilde \varphi}(\xi)$, such that ${\tilde \varphi}(\xi)\to
\varphi_+$ as $\xi\to+\infty$ and ${\tilde \varphi}(\xi)\to \varphi_-$
as $\xi\to-\infty$ and $\tilde \varphi'(\xi)$ is an even function;

(d) the solution ${\tilde \varphi}(\xi)$ can be continued into the
complex plane and the closest to the real axis singularities of
${\tilde \varphi}(\xi)$ in the upper-half plane form a pair which
is symmetric with respect to imaginary axis,
\begin{eqnarray}
z_{\pm} = \pm \alpha + i \beta,\quad \alpha, \beta> 0.\label{ConjCond}
\end{eqnarray}
\medskip

Then one can expect {\it an infinite sequence of parameter values
$\lambda=\lambda_n$, $n=1,2,\ldots$}, such that for each of them
Eq.~(\ref{MainEq}) has a kink solution and this sequence obeys the
asymptotics
\begin{eqnarray}
k(\lambda_n)\sim\left(n\pi+\theta_0\right)/\alpha, \label{AsRel}
\end{eqnarray}
where $k(\lambda)$ is the real root of Eq.~(\ref{DisRelKink}),
$\alpha$ is the modulus of real part of singularity $z_{\pm}$ and
$\theta_0$ is a constant.}

Up to the moment we have no rigorous proof of the Conjecture stated
above (some heuristic arguments are presented in \cite{arXiv}). At the
same time the asymptotic formula (\ref{AsRel}) has been strongly
confirmed numerically for various examples of operator $G_\lambda$ and
the function $F(\varphi)$. Note that the type of the singularity of
${\tilde \varphi}(\xi)$ is not specified, it can be pole, logarithmic
or transcendental branching point etc.\medskip

Let us show that the Conjecture gives remarkable good results for
Eq.~(\ref{eq-exp-2}).\medskip

{\bf 1.} Let us check the points (a) --- (d) for Eq.~(\ref{eq-exp-2}).
The equilibrium states are $\varphi_-=0$ and $\varphi_+=2\pi$. The
operator $G_\lambda$ has the symbol
\begin{eqnarray*}
\hat{G}_\lambda=-v^2k^2+\frac{k^2}{1+\lambda^2k^2}
\label{SymbolEx}
\end{eqnarray*}
and $G_0=(1-v^2)d^2\varphi/d\xi^2$, i.e $\Omega=\sqrt{1-v^2}$. The
relation (\ref{DisRelKink}) has the form (\ref{DispRel}). The positive
root $k(\lambda)$ is unique and has the asymptotics
\begin{eqnarray*}
k(\lambda)\simeq \frac{\sqrt{1-v^2}}{\lambda v}, \quad {\rm as} \quad
\lambda\to 0.
\end{eqnarray*}
The $2\pi$-kink solution of the equation $(1-v^2)
d^2\varphi/d\xi^2=F(\varphi)$ has the form~(\ref{2SG_Sol}). If
$A>0$ then the closest to the real axis singularities of
(\ref{2SG_Sol}) in the upper complex half-plane are
$z_\pm=\pm\alpha+i\beta$ where
\begin{eqnarray*}
\alpha=\frac{\sqrt{1-v^2}}{2 \sqrt{1+4A}} {\rm arccosh}(1 +
8A),\quad \beta =\frac{\pi \sqrt{1-v^2}}{2\sqrt{1+4A}}.
\end{eqnarray*}
Then, according to Conjecture one can expect that for $v$ fixed
\begin{equation}
\lambda_n \sim \frac{(1-v^2){\rm arccosh}(1 + 8A)}{v
\sqrt{1+4A}((2n-1)\pi+\tilde{\theta}_0)}, \quad
 n\to\infty, \label{Ex2As}
\end{equation}
where $\tilde{\theta}_0=2\theta_0-\pi$, see (\ref{AsRel}).  The Table
\ref{T1} shows the values of $\lambda_n$ computed numerically in
comparison with ones calculated by the formula (\ref{Ex2As}) with
$\tilde{\theta}_0=0$ (the arguments about this choice of
$\tilde{\theta}_0$ can be found in \cite{arXiv}). It follows from
Table~\ref{T1} that the correspondence between numerical and
asymptotical values is quite good for large enough $n$. However, for
$n$ large the function $R_\lambda(v)$ (mentioned in Sect.\ref{NR_Main})
is small, and there appears a difficulty of localizing its zeroes. This
explains empty entries in the Table {\ref{T1} for $n=5,6$ and
$v=0.5$.\medskip

{\bf 2.} If $A>0$ for Eq.~(\ref{4thOrdScale}) the points (a) --- (d)
are also fulfilled. The equilibrium states are $\varphi_-=0$ and
$\varphi_+=2\pi$. The operator $G_\delta$ has the symbol
\begin{eqnarray*}
\hat{G}_\delta=k^2+\delta^2 k^4 \label{SymbolEx4}
\end{eqnarray*}
and $G_0=d^2\varphi/d\eta^2$. The relation (\ref{DisRelKink})
reads
\begin{eqnarray}
k^2+\delta^2k^4=1+4A. \label{DR4}
\end{eqnarray}
Eq.~(\ref{DR4}) has unique positive root $k(\delta)\sim 1/\delta$,
$\delta\to 0$. The $2\pi$-kink solution of the equation
$d^2\varphi/d\eta^2=F(\varphi)$ is
\begin{eqnarray*}
\tilde{\varphi}(\eta)=\pi + 2\arctan\left(\frac1{\sqrt{1+4A}}
\sinh\left(\sqrt{1+4A}\eta\right)\right)
\end{eqnarray*}
and its closest to the real axis singularities in the upper
complex half-plane for $A>0$ are $z_\pm=\pm\alpha+i\beta$ where
\begin{eqnarray*}
\alpha=\frac{{\rm arccosh}(1 + 8A)}{2 \sqrt{1+4A}} ,\quad \beta
=\frac{\pi }{2\sqrt{1+4A}}.
\end{eqnarray*}
Then, according to the Conjecture one can expect that
\begin{equation}
\delta_n \sim \frac{{\rm arccosh}(1 + 8A)}{
\sqrt{1+4A}((2n-1)\pi+\tilde{\theta}_0)},\quad
n=1,2,\ldots\label{Ex2As4D}
\end{equation}
The Table \ref{T2} presents the values of $\delta_n$ computed both,
numerically and by the formula (\ref{Ex2As4D}) with
$\tilde{\theta}_0=0$. It follows from Table \ref{T2} that the
correspondence between numerical and asymptotical values is also good
for large enough $n$. Also for $n=6,7,8$ and $A=1$ the accuracy of the
numerical computation is insufficient and the results are not
exposed.\medskip

\begin{table}
  \begin{tabular}{|l|l|c|c||l|l|c|c|}  \hline
 $v$& $n$ &Asympt. {$\lambda_n$}& Calcul.
{$\lambda_n$}&$v$ & $n$ &  Asympt. {$\lambda_n$}  & Calcul.
$\lambda_n$
 \\ \hline
 $0.1$ &1 & 3.3885 & 0.9116 & $0.5$ &1 & 0.5134 & 0.3751 \\ \hline
  &2 & 1.1295 & 0.6809 &&2 & 0.1711 & 0.1698  \\ \hline
  &3 & 0.6777 & 0.5319 &&3 & 0.1026 & 0.1043  \\ \hline
  &4 & 0.4840 & 0.4311 &&4 & 0.0733 & 0.0745  \\ \hline
  &5 & 0.3765 & 0.3584 &&5 &   -    &    -      \\ \hline
  &6 & 0.3080 & 0.3041 &&6 &   -    &    -       \\ \hline
  \end{tabular}
  \caption {Comparison of asymptotical and numerical values of $\lambda$ when $2\pi$-kink solutions exist
  for Eq.~(\ref{eq-exp-2}): $A=1/8$. In the cases $n=5,6$ and $v=0.5$ the accuracy of the numerical computation
  is insufficient and the results are not exposed.}\label{T1}
\end{table}

\begin{table}
  \begin{tabular}{|l|l|c|c||l|l|c|c|}  \hline
 $A$& $n$ &Asympt. {$\delta_n$}& Calcul.
{$\delta_n$}&$A$ & $n$ &  Asympt. {$\delta_n$}  & Calcul.
$\delta_n$
 \\ \hline
 $1$  &1 & 0.4110 & 0.3149 & 10& 1& 0.2529 & 0.1320   \\ \hline
  &2 & 0.1370 & 0.1350 & & 2 & 0.0843 & 0.0664 \\ \hline
  &3 & 0.0822 & 0.0823 & & 3 & 0.0506 & 0.0486  \\ \hline
  &4 & 0.0587 & 0.0588 & & 4 & 0.0361 & 0.0364  \\ \hline
  &5 & 0.0456 & 0.0457 & & 5 & 0.0281 & 0.0283  \\ \hline
  &6 & -      & -      & & 6 & 0.0230 & 0.0231 \\ \hline
  &7 & -      & -      & & 7 & 0.0195 & 0.0195 \\ \hline
  &8 & -      & -      & & 8 & 0.0169 & 0.0169 \\ \hline
  \end{tabular}
  \caption {Comparison of asymptotical and numerical values of $\delta$ when $2\pi$-kink solutions exist
  for Eq.~(\ref{4thOrdScale}): $A=1$ and $A=10$. In the cases $n=6,7,8$ and $A=1$ the accuracy of the numerical computation
  is insufficient, and the results are not exposed.}\label{T2}
\end{table}

Since the correspondence between the numerical and asymptotical results
is good we assume that the Conjecture is valid and there exists
infinitely many branches $v_n(\lambda)$ for $A>0$. Let us note that in
the interval $-1/4<A<0$ there also exists $2\pi$-kink solution
(\ref{2SG_Sol}) of Eq.~(\ref{2SGTrW}) but its closest to the real axis
singularities are situated on imaginary axis, therefore the conditions
(\ref{ConjCond}) of the Conjecture do not hold. Numerical study does
not reveal for $-1/4<A<0$ {\it any} root of the function
$R_\lambda(v)$. Consequently, we conclude that no $2\pi$-kinks with
nonzero velocities exist for $A<0$ in the nonlocal model.

\section{Propagation of fluxons: numerical experiments}\label{Prop}

In this section we consider the evolution  governed by the nonlocal
equation (\ref{eq-exp-2}). We show that the model supports the
propagation of  $2\pi$-kinks of Sect. \ref{NR_Main} in radiationless
regime with the velocities $v_n(\lambda)$. Moreover, we argue that this
regime of propagation {\it is asymptotical} for some class of kink-like
excitations. Below all the results are shown for $\lambda = 0.3$ and
$A= 1/8$. For this case the first and second discrete velocities for
radiationless propagation are $v_1\approx 0.5831$ and $v_2\approx
0.3213$. The corresponding values of energy integral are ${\cal
W}_1\approx 10.9490$ and ${\cal W}_2\approx7.9399$. The profiles of
$2\pi$-kinks corresponding to the first and second discrete velocities
were found by numerical solution of (\ref{Eq1})~---~(\ref{Eq2}). We
denote them by $\varphi_1(\xi)$ and $\varphi_2(\xi)$.\medskip

1. First, $2\pi$-kink $\varphi_1(\xi)$ was furnished at the initial
moment its {\it natural} velocity $v=v_1$.  The evolution of this
entity is shown in Fig.~\ref{front_kink}, panel A. It follows from this
figure that the velocity is conserved and no radiation appears.

\begin{figure} [h]
\centerline{\includegraphics[scale=0.4]{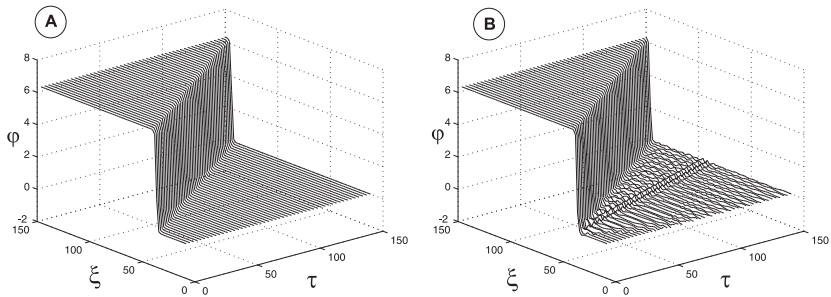}}
 \caption{$2\pi$-kink propagation for $A= 1/8$, $\lambda = 0.3$. (A)
 Radiationless propagation of the kink $\varphi_1(\xi)$ with the first discrete velocity $v=v_1$; (B) propagation
of the kink of the same shape as in (A) but supplied with velocity
$v=0.9$.} \label{front_kink}
\end{figure}

Then the same $2\pi$-kink $\varphi_1(\xi)$ was furnished a velocity $v=
0.9>v_1$. The energy integral in this case is ${\cal
W}\approx12.9662>{\cal W}_1$. Fig.~\ref{front_kink}, panel B and
Fig.~\ref{velocity_for_front} show that the travelling kink slows down
until its natural velocity $v_1$. Extra energy has been radiated. The
similar phenomena for the model $\varphi^4-\varphi^6$ was reported on
in \cite{AlfMed11}.

\begin{figure} [h]
\centerline{\includegraphics[scale=0.5]{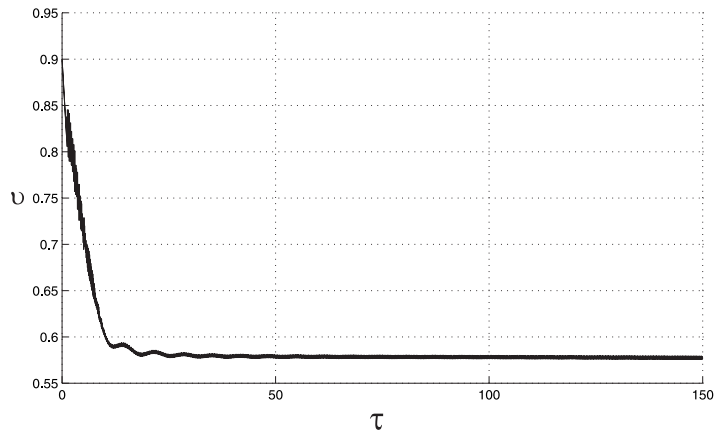}}
 \caption{Velocity of the kink front center (the point where $\varphi =\pi$) corresponding
to the propagation in the panel B in Fig.~\ref{front_kink}.}
\label{velocity_for_front}
\end{figure}

2. In the next series of experiments we considered the evolution of
kink-like excitation of the form
\begin{eqnarray}
\varphi=4\arctan{\left(\exp{(\frac{\gamma
\xi}{\sqrt{1-v^2}})}\right)}.\label{SGkink}
\end{eqnarray}
Here $\gamma$ is a coefficient describing the slope of the kink
front.\medskip

(i) Fig.~\ref{first_exc} shows the propagation of this kink-like
excitation for $\gamma=0.5$ supplied at the initial moment $\tau=0$
with a velocity $v=0.99$. This velocity is greater than the first
discrete velocity $v_1$ of radiationless $2\pi$-kink propagation. The
energy integral in this case is ${\cal W} \approx 26.2718$ which is also greater than ${\cal
W}_1$.

The profiles of the kink-like excitation (\ref{SGkink}) and $2\pi$-kink
solution $\varphi_1(\xi)$ are depicted in Fig.~\ref{first_exc}, panel A.
Then, the profiles were compared at the moment $\tau=120$
(Fig.~\ref{first_exc} panel B). One can observe that they match each
other quite well. The decreasing of the velocity for the kink-like
excitation is shown in Fig.~\ref{first_exc}, panel C. Note that the
velocity in this case changes smoothly, without large oscillations.

\begin{figure} [t]
\centerline{\includegraphics[scale=0.3]{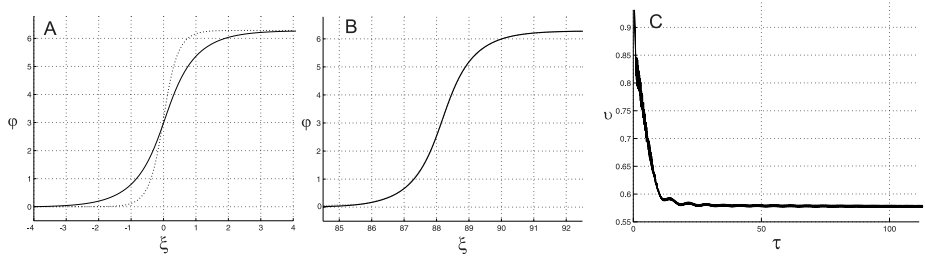}}
 \caption{Propagation of kink-like excitation (\ref{SGkink}), $\gamma=0.5$, $v=0.99$.
  In panels A-B: solid line - the profile of $2\pi$-kink solution of Eq.~(\ref{TrW}) for $A= 1/8$, $\lambda = 0.3$ corresponding
  to the first discrete velocity $v_1\approx0.5831$; dotted line -  the profile of the kink-like excitation.
  (A)  $\tau=0$;   (B)  $\tau=120$; (C) velocity of the  kink-like excitation
 (measured as velocity of the point $\varphi =\pi$ on the kink front). } \label{first_exc}
\end{figure}

\begin{figure} [h]
\centerline{\includegraphics[scale=0.3]{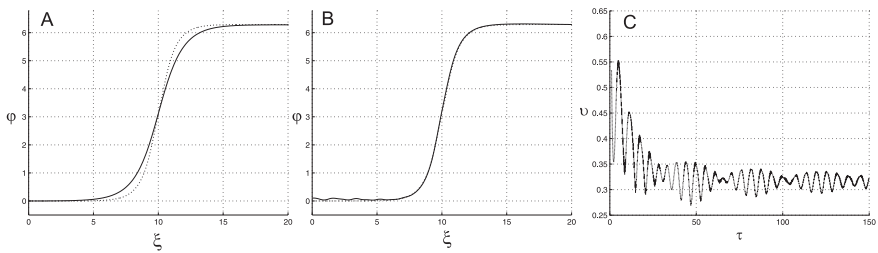}}
 \caption{Propagation of kink-like excitation of Eq.~(\ref{SGkink}), $\gamma=0.5$, $v=0.8$.
 In panels A-B: solid line - $2\pi$-kink solution for $A= 1/8$, $\lambda = 0.3$ corresponding to
 the second discrete velocity $v_2\approx0.3213$ in Eq.~(\ref{TrW}); dotted
 line - the profile of kink-like excitation. (A)  $\tau=0$;   (B)  $\tau=150$;
 (C) velocity of the  kink-like excitation.} \label{sec_exc}
\end{figure}

(ii) The kink-like excitation (\ref{SGkink}) with $\gamma=0.5$  was
furnished initially a velocity $v=0.8$. This velocity is greater than
the first discrete velocity $v_1$ of $2\pi$-kink solution
$\varphi_1(\xi)$. However, the value of energy integral of the
excitation ${\cal W}\approx 9.4018 $ lies between the first and second
discrete energy values ${\cal W}_1$ and ${\cal W}_2$. Numerical
simulation shows  (see Fig.~\ref{sec_exc} panel A) that in this case the
velocity falls down to the second discrete velocity $v_2$.
Fig.~\ref{sec_exc}, panels A and B, show the difference between the
profiles of kink-like  excitation and and $2\pi$-kink solution
$\varphi_1(\xi)$ at the moment $\tau=0$ and $\tau=150$. Again the
profiles in Fig.~\ref{sec_exc} panel B match well each other. In this
case the velocity of kink-like excitation exhibits considerable
oscillations (Fig.~\ref{sec_exc}, panel C).

So, the numerical experiments show that the phenomenon of quantization
of $2\pi$-kink velocities is essential for dynamics of nonlocal model.
In the process of evolution a kink-like excitation ``adjusts'' its
shape and velocity approaching to one of $2\pi$-kinks which propagates
in radiationless regime. The energy $\cal W$ of initial state is
crucial for selection of the velocity of such propagation.

\section{Discussion}\label{Disc}

To summarize, we have studied the properties of vortices in
stacked Josephson structures. Two assumptions have been made: (i) the
electrodynamics of Josephson structure is {\it nonlocal} and (ii) the
current-phase relation is represented by {\it two sine harmonics},
instead of one sine harmonic for the sine-Gordon case. The main
equation for the model is the nonlocal double sine-Gordon (NDSG)
equation. This equation depends on two governing parameters: $\lambda$
which measures the ``strength'' of nonlocality and $A$ which is the
amplitude of the second harmonic in current-phase relation. The limit
$\lambda=0$ corresponds to traditional (local) double sine-Gordon
equation, whereas the limit $A=0$ corresponds to nonlocal sine-Gordon
equation which has been discussed in many studies \cite{AAM,AEKM,AEL}.

The simplest Josephson vortex (fluxon) corresponds to $2\pi$-kink
solution of NDSG equation. We have found that the velocity for
radiationless motion of this $2\pi$-kink cannot be arbitrary. Our
study allows to suppose that there are infinitely many branches
$v_n(\lambda)$ of the velocities of the kink. Some asymptotic
properties of families $v_n(\lambda)$ are also described.

Numerical simulations of vortex propagation reveal the following
picture. Let the nonlocality parameter $\lambda$ be fixed. Then there
exist the set of kink velocities $v_1(\lambda)>v_2(\lambda)>\ldots$ and
the set of energy values $W_1(\lambda)>W_2(\lambda)>\ldots$,
corresponding to propagation of kink with these velocities. Kink-like
excitation launched with velocity greater than $v_1(\lambda)$ with the
energy greater than $W_1(\lambda)$ radiates and slows down to the
velocity $v_1(\lambda)$. If the initial energy of kink-like excitation
lies between $W_1(\lambda)$ and $W_2(\lambda)$, it radiates and slows
down to the velocity $v_2(\lambda)$. So, the process of evolution does
``feel'' the selected values of $2\pi$-kink radiationless propagation.

It is important to stress that there are significant differences
of NDSG equation from both, double sine-Gordon equation and
nonlocal sine-Gordon equation. Specifically
\begin{itemize}
\item[(i)] In the local case $\lambda=0$ the $2\pi$-kink can
have arbitrary velocity $v^2<1$. In the case of NDSG equation the
spectrum of kink velocities is discrete;
\item[(ii)] In the case of nonlocal sine-Gordon equation $2\pi$-kink can have zero velocity
only, therefore $2\pi$-kinks cannot propagate. At the same time there
exist many kinks with higher topological charge (i.e. $4\pi$-,
$6\pi$-kinks etc) and each of them corresponds to its own velocity of
propagation. In the case of NDSG model the phenomenon of velocities
quantization takes place for $2\pi$-kinks. Moreover, the profiles of
$2\pi$-kinks corresponding to different velocities are very similar and
differ mainly in asymptotics of the ``tails''.
\end{itemize}

\section{Acknowledgement}\label{ack}

Authors are grateful to Prof. D.~Pelinovsky for useful discussion.
Authors are grateful to Prof. M. Yu. Kupriyanov who communicated
information on articles \cite{Gol-2005, Gol-2004}.  The work of GLA and
EVM was supported by grant of Russian Foundation for Basic Research
13-01-00199.

\appendix


\section{Josephson layered structure with thin layers: derivation of nonlocal double sine-Gordon equation}\label{Deriv}

Consider a Josephson structure consisting of alternating
superconducting layers of thickness $2L$, strips
$(2n-1)d+2(n-1)L<y<(2n-1)d+2nL$, and nonsuperconducting (tunnel) layers
of thickness $2d$, strips $(2n-1)d+2nL<y<(2n+1)d+2nL$,
$n\in\mathbb{Z}$. Assume that the phase differences for the
superconducting order parameters of the electrodes on different sides
of the all tunnel layers are the same and are equal to $\varphi(x,t)$.
Assume also that the magnetic field $H_J(x,t)$ is also the same in all
tunnel layers. Then the magnetic field in $n$-th superconducting layer
obeys the London equation
\begin{eqnarray}
\frac{\partial^2 \,}{\partial x^2}H(x,y,t)+\frac{\partial^2
\,}{\partial y^2}H(x,y,t)=\frac1{\lambda^2_L}
H(x,y,t), \label{London}
\end{eqnarray}
where $\lambda_L$ is the London penetration length. The boundary
conditions for Eq.~(\ref{London}) are
\begin{eqnarray*}
H(x,y_n-d-2L,t)=H(x,y_n-d,t)=H_J(x,t),\quad y_n\equiv 2n(d+L).
\end{eqnarray*}

Assume that the superconducting layers are thin and the following
condition holds
\begin{eqnarray}
2L\sqrt{\lambda_L^{-2}+k^2} \ll 1, \label{cond-thin}
\end{eqnarray}
where $k$ is the inverse characteristic spatial scale of the
variation of the phase difference along the layered structure.
Then the solution of (\ref{London}) can be approximated by the
following formula
\begin{eqnarray}
H(x,y,t)\simeq\left[1+\frac12(y-y_n+d+2L)(y-y_n+d)\left(-\frac{\partial^2
\,}{\partial
x^2}+\frac1{\lambda_L^2}\right)\right]H_J(x,t). \label{H}
\end{eqnarray}
Taking into account similar relation for $(n+1)$-th superconducting
layer and making use of relation
\begin{eqnarray*}
{\bf E}=\frac{\lambda^2_L}c {\rm rot}~\frac{\partial{\bf
H}}{\partial t}
\end{eqnarray*}
we arrive at the following formula for the jump of tangential
component of electric field across $n$-th tunnel layer
\begin{eqnarray}
E_x(x,y_n+d,t)-E_x(x,y_n-d,t)=-\frac{2\lambda_L^2L}{c}\left(-\frac{\partial^2
\,}{\partial x^2}+\frac1{\lambda_L^2}\right)\frac{\partial
H_J(x,t)}{\partial t}, \label{E-jump}
\end{eqnarray}
where $c$ is the speed of light in vacuum. Taking into account
(\ref{E-jump}), the relation between the normal component of
electric field ${\bf E}$ in the tunnel layer with the phase difference
\begin{eqnarray*}
E_y(x,t)=\frac{\phi_0}{4\pi c d}\frac{\partial\varphi\,}{\partial
t}(x,t)
\end{eqnarray*}
and $z$-component of Maxwell equation
\begin{eqnarray*}
{\rm rot}~{\bf E}=-\frac1c\frac{\partial{\bf H}}{\partial t}
\end{eqnarray*}
we arrive at the expression for magnetic field in tunnel layer
\begin{eqnarray}
H_J(x,t)=-\frac{\phi_0\lambda_{eff}}{8\pi\lambda_L^2L}\int~dx'\exp\left\{-\frac{|x-x'|}{\lambda_{eff}}\right\}\frac{\partial\varphi(x',t)}{\partial
x'},\label{H_J}
\end{eqnarray}
Here $\phi_0$ is the magnetic flux quantum and
\begin{eqnarray*}
\lambda_{eff}\equiv\frac{\lambda_L}{\sqrt{1+d/L}}.
\end{eqnarray*}
Then the density of superconducting current on the boundary
between superconducting and nonsuperconducting layers is
\begin{eqnarray*}
-\frac{c}{4\pi}\frac{\partial{H_J(x,t)}}{\partial x}.
\end{eqnarray*}
The matching condition for the current on this boundary yields
\begin{eqnarray}
j(\varphi)+\frac{\epsilon\phi_0}{16\pi^2 cd}\frac{\partial^2
\varphi}{\partial
t^2}=\frac{c\phi_0\lambda_{eff}}{32\pi^2\lambda_L^2
L}\frac{\partial\,}{\partial
x}\int~dx'\exp\left\{-\frac{|x-x'|}{\lambda_{eff}}\right\}\frac{\partial\varphi(x',t)}{\partial
x'},\label{FinalEq}
\end{eqnarray}
where $j(\varphi)$ is density of Josephson supercurrent, $\epsilon$ is the permittivity
of the tunnel layer (cf. with Eq.~(5.6) of article \cite{AOSU}).

 For the energy of one period of the
layered structure, (i.e. for the area $y_n-d-2L<y<y_n+d$) per unit
of $Oz$ one has
\begin{eqnarray*}
W&=&\frac{\phi_0}{2\pi c}\int~dx\left[\int_0^\varphi
j(\varphi')~d\varphi'+\frac{\phi_0\varepsilon}{32\pi^2c
d}\left(\frac{\partial \varphi}{\partial
t}\right)^2\right]+\\[2mm]
&+&\frac{\phi_0^2\lambda_{eff}}{128\pi^3\lambda_L^2
L}\int\int~dx_1~dx_2\exp\left\{-\frac{|x_1-x_2|}{\lambda_{eff}}\right\}\frac{\partial\varphi(x_1,t)}{\partial
x_1}\frac{\partial\varphi(x_2,t)}{\partial x_2}.
\end{eqnarray*}

If $j(\varphi)$ is defined by the formula $j_c\sin\varphi + j_2\sin2\varphi$
(cf. formula (\ref{DoubleHarm}) for supercurrent), the energy takes the following
form:
\small
\begin{eqnarray}
&&W=\frac{\phi_0 j_c}{2\pi c}\left\{\int~dx\left[1-\cos\varphi+A(1-\cos
2\varphi)+\frac{1}{2\omega_J^2}\left(\frac{\partial \varphi}{\partial
t}\right)^2\right]\right.+\nonumber\\[2mm]
&+&\left.\frac{(\lambda_L+d)\lambda_J^2}{4\lambda_L\sqrt{(L+d)L}}\int\int~dx_1~dx_2\exp\left\{-\frac{|x_1-x_2|}{\lambda_{eff}}\right\}\frac{\partial\varphi(x_1,t)}{\partial
x_1}\frac{\partial\varphi(x_2,t)}{\partial
x_2}\right\},\label{EnIntegral}
\end{eqnarray}
\normalsize where $\omega_J\equiv 4\pi\sqrt{c j_c d/\phi_0\epsilon}$ is
the Josephson plasma frequency and
\begin{eqnarray*}
\lambda_J\equiv \frac1{4\pi}\sqrt{\frac{\phi_0
c}{j_c(\lambda_L+d)}}
\end{eqnarray*}
is the Josephson length.


\end{document}